\documentclass[preprint]{article}

\usepackage[english]{babel}

\usepackage[letterpaper,top=2cm,bottom=2cm,left=3cm,right=3cm,marginparwidth=1.75cm]{geometry}

\usepackage{natbib}
\usepackage{amsmath}
\usepackage{graphicx}
\usepackage{subcaption}
\usepackage{caption}
\usepackage[colorlinks=true, allcolors=blue]{hyperref}

\title{Agent-based model of information diffusion\\in the limit order book trading}

\author{Mateusz Wilinski, Juho Kanniainen}

\begin{document}
\maketitle

\begin{abstract}
There are multiple explanations for stylized facts in high-frequency trading, including adaptive and informed agents, many of which have been studied through agent-based models. This paper investigates an alternative explanation by examining whether, and under what circumstances, interactions between traders placing limit order book messages can reproduce stylized facts, and what forms of interaction are required. While the agent-based modeling literature has introduced interconnected agents on networks, little attention has been paid to whether specific trading network topologies can generate stylized facts in limit order book markets. In our model, agents are strictly zero-intelligence, with no fundamental knowledge or chartist-like strategies, so that the role of network topology can be isolated. We find that scale-free connectivity between agents reproduces stylized facts observed in markets, whereas no-interaction does not. Our experiments show that regular lattices and Erdős–Rényi networks are not significantly different from the no-interaction baseline. Thus, we provide a completely new, potentially complementary, explanation for the emergence of stylized facts.
\end{abstract}

\section{Introduction}

Modeling financial markets and other economic systems with a bottom-up approach, called agent-based modeling, has been present in the literature for a while \cite{arthur2006out}.
One of the first uses of the term \textit{Agent} in the modeling context is attributed to an article on economic theory \cite{holland1991artificial}.
Interest in this type of out-of-equilibrium economic models has grown during the 2008 financial crisis \cite{klimek2015bail}.
A successful example of applying agent-based modeling in finance is the zero-intelligence approach to the limit order book modeling \cite{gode1993allocative}.
It was specifically useful in understanding the shape of the price impact function \cite{farmer2005predictive} and showing how it can emerge from the double auction mechanism itself.

Agent-based modeling has vast applications, including epidemics \cite{pastor2015epidemic}, systemic risk \cite{haldane2011systemic}, opinion dynamics \cite{acemoglu2011opinion}, game theory \cite{axelrod1997complexity}, and many others.
In most fields where agent-based modeling is used, one of the crucial components is the network describing the interactions between the agents.
When it comes to stock market, many existing agent-based models assume that the interaction is mediated through the market.
If, for example, agents' strategies are a function of price formation, their behavior depends on other market participants, but the interaction is not direct \cite{staccioli2021agent}.
Such view is in line with a vision of a trader whose activity is governed fully by the price activity observed on the screen in front of him.
However, the reality is more complex.
On one hand, trading patterns suggest existence of information networks connecting traders \cite{baltakiene2021identification}.
Moreover, the market is becoming networked even in the high frequency domain \cite{musciotto2021high}.
On the other hand, we observe an increased impact of social media activities on the stock market \cite{ranco2015effects}.
The latter is reinforced by influential figures, who reach large groups of people and affect, sometimes on purpose, sometimes unintentionally, market prices \cite{shahzad2022price}.
Even if one believes that in the past the interaction between market participants was only mediated through the market, this is no longer true.
While the agent-based modeling literature has introduced interconnected agents on networks, little attention has been paid to whether specific trading network topologies can generate stylized facts in Limit Order Book markets.

Initially work on interacting agents did not focus on local connectivity, but assumed full observation of all agents.
Influential model in this setting was proposed in \cite{brock1998heterogeneous}, where agents modified their strategy based on past profitability of all investors.
Similar direction was taken in \cite{lux2000volatility}, but with more focus on statistical properties of the resulting price.
In both cases agents are supposed to base their strategy on the fundamental price (fundamentalists), or on technical analysis (chartists), or a mix of the two.
In \cite{cont2000herd} agents simply decide the direction of trades, and a random network is introduced in order to split them into cooperating clusters.
This gives rise to heavy tails of price returns, which are computed using demand excess.
An Ising-like model on a lattice, with an additional budget constraint, was used in \cite{iori2002microsimulation}.
Interaction through a lattice in the context of fundamentalists and chartists was proposed in \cite{horst2005financial}.
Early work with multiple network structures in the context of asset price dynamics was \cite{alfarano2009network}, but it was limited to the N-dependence problem.
A more detailed analysis of stability as a function of structure based on the Watts-Strogatz network \cite{watts1998collective} was done in \cite{panchenko2013asset}.
Important direction, from our perspective, was to move towards order-driven mechanism, instead of using \textit{ad hoc} price models based on supply and demand \cite{lebaron2007long}.
The order-driven market, together with the dynamic network, was shown in \cite{tedeschi2009role} and further developed in \cite{tedeschi2012herding}.
In these two articles basic agent behavior was of zero-intelligence type, but it was influenced by neighbors.
More importantly, some statistical properties of price dynamics were reconstructed, but only with additional conditions on network evolution and agents behavior.
Some more recent work includes \cite{huang2023dynamic, hatcher2024communication}.
All the described so far models were of the opinion dynamics type, but a recent model from \cite{gong2023impacts} includes spreading as a mechanism for changing strategy.
However, it is still in the paradigm of chartists competing with fundamentalists, and price is modeled as a linear function of demand.
A good, and recent review of networked models for asset price dynamics can be found in \cite{hatcher2024communication}.

In this work we focus on the interplay between interaction network and price formation.
To this end, we limit all the other mechanisms that can lead to non-trivial statistical price properties.
Firstly, we use an order-driven double auction mechanism, in order to simulate realistic price dynamics.
Second, we introduce a simple, but novel agent-based model for limit order book, where agents exchange information through spreading on a predefined network.
Agents are of zero-intelligence type with no fundamental knowledge, or chartists-like strategies, so that the
role of network topology can be isolated.
The interaction mechanism exhibits similarity to classical spreading models on networks \cite{newman2011structure}.
This is justified by an empirical study, which suggests that spreading (epidemic) models for social influence can predict price returns \cite{shive2010epidemic}.
The closest spreading mechanism, in the context of trading agents, we found is described in \cite{toth2015equity}, but it is limited to trade direction and not analyzed in the context of different connectivity.
To the best of our knowledge, all the previous models in this field, including the one from \cite{toth2015equity}, were of sequential type.
Our model, on the contrary, allows for non-trivial waiting times, and as such, can be used to test the statistical properties of the latter.

Full description of the model is available in Section \ref{sec:abm}.
In Section \ref{sec:results} we test the model against well established statistical properties of the stock market, known from the market microstructure literature \cite{bouchaud2018agent}.
Finally, we discuss the results in Section \ref{sec:discussion}.
The main conclusion of our work is that simple spreading mechanism, coupled with appropriate network structure, can reproduce multiple \textit{stylized facts} known from empirical, financial data.

\section{Interacting Agent-Based Model}\label{sec:abm}

\subsection{Limit Order Book}

In our simulation we use an artificial continuous double auction market.
Actions available to the agents are limited to limit orders, market orders and cancellations.
Limit orders can have a maximum execution price, for buy orders, and minimum execution price, for sell orders.
Market orders are immediately executed at the best available price.
If a limit order does not get entirely executed (not enough volume available at the execution limit of the order), then the order will be stored in a limit order book \cite{gould2013limit}.
Orders are stored and matched using the price-time priority \citep{preis2011price}.
This means that buy orders with higher price are prioritized on the bid side, while ask side prioritize sell orders with lower price.

\begin{figure}[h!]
\centering
\includegraphics[width=0.50\textwidth]{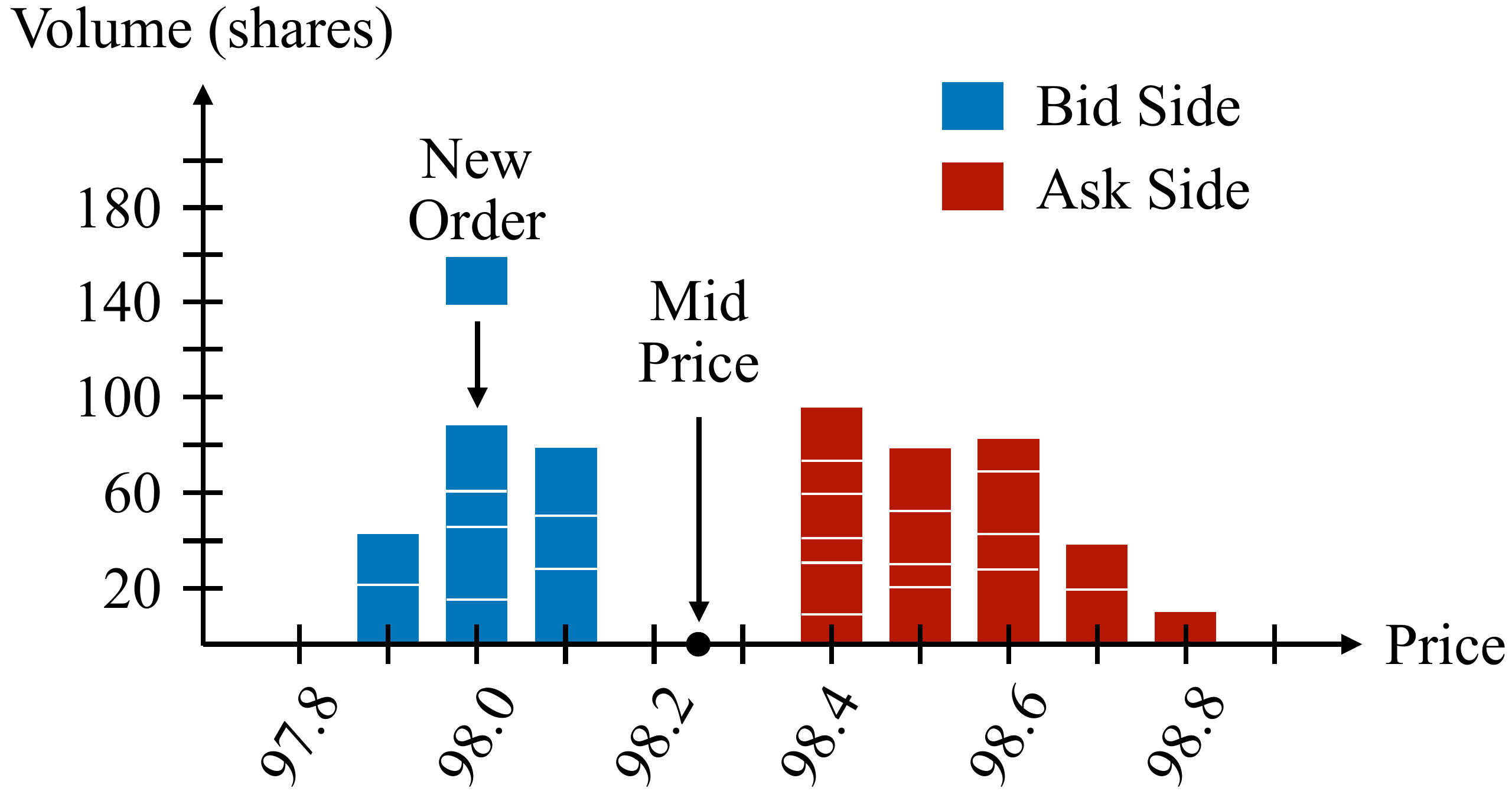}
\caption{A snapshot of a limit order book with an incoming 20 shares buy order at a limit price of 98.0.
The best buy offer is at 98.1 and the best sell offer is at 98.4, which result in a 98.25 mid price and a 0.3 spread.
Tick size in this experiment is equal to 0.1.
\label{fig:lob}}
\end{figure}

A snapshot of a limit order book is shown in Fig. \ref{fig:lob}.
Limit order for 20 shares is send to the bid side at the price of 98.0.
The best bid offer $b_t$ is equal to 98.1, while the best ask offer $a_t$ is equal to 98.4.
The mid-price is the average over the two $p_t = (a_t + b_t) / 2$.

The limit order book mechanism, together with all the experiments used in this work, is available at \citep{wilinski2025simulator}, where all the technical details can be found.
The simulation is driven by messages sent between the agents and the engine, rather than based on discrete time steps.

\subsection{No interaction behavior}\label{sec:no-interaction}

Agents used in the simulation are of the zero-intelligence type.
They can be seen as irrational investors whose decisions are based on speculation \citep{shleifer1990noise}.
Their behavior in the network model proposed in this work can be divided into two types: individual, and interaction-related.
In our simplified model the former is not directly affected by the latter.
To make it more clear, let us start by describing the actions of the agents when connections are absent.
In such case, each agent has three available actions: placing a market order, placing a limit order, or canceling an active (stored in the limit order book) order.
In all three cases, there is a similar mechanism when it comes to timing of agent's actions.
The agent appears at the market at times $t_1, t_2, \dots$, where each $t_i$ is chosen randomly such that $\delta t = t_i - t_{i-1}$ is described by an exponential distribution $\lambda e^{\lambda \delta t}$, and performs a given action.
For market orders, it means placing the order with randomly picked direction.
Volume is drawn from a Gaussian distribution and rounded to the closest integer (but never below one share).
For limit orders, in addition to direction and volume, agent needs to pick a price.
He does it by choosing a number from a log-normal distribution with mean given by the current mid-price $p_t$ and a fixed standard deviation.
Note that the order's price is independent of the chosen side and is rounded to the closest tick.
Finally, the cancellation action is performed by randomly choosing one of the active orders (using a uniform distribution) placed by the agent, and deleting it from the limit order book.
The arrival times for these three types of actions are drawn independently.
This means that the agent has three (potentially) distinct parameters describing the exponential distribution of waiting times ($\lambda_m$ for market orders, $\lambda_l$ for limit orders, $\lambda_c$ for cancellations).
An agent is also parametrized by mean order volume $m_s$, the standard deviation of the order volume $d_s$, and the standard deviation of selected price $d_p$.
In our simulations all of these parameters are fixed and equal for all the agents.
The price selection can be seen in terms of choosing a logarithmic price difference, i.e. we choose a number from Gaussian distribution centered at zero, and multiplying the current mid-price accordingly.
This way, all the actions are independent not only from each other, but also from the market dynamics itself.
In other words, we could choose each agent's behavior before even running the simulation.
Therefore, it cannot be affected by the interaction-related actions, which we will describe in the next subsection.
From now on, we will refer to the non-interaction behavior described above as \textit{source} actions.

\subsection{Information spreading}

The interactions between the agents is modeled as information spreading and its mechanism is similar to classical spreading models on networks.
Once an agent makes a decision to place a source order, he shares this information with his neighbors.
In our study, the network of connections between the agents is unweighted, undirected, and constant in time.
Each neighbor decides, with a fixed probability $q$, whether to follow the given suggestion.
If the decision is positive, the neighbor places a new order.
The time between the source order and the follow-up order from the neighbor is exponentially distributed, with $\lambda_f$ rate fixed and equal for all the agents.
Neighbor's order is of the same type (market or limit type) as the source order.
The direction of neighbor's order is the same as the direction of the source order, but the size is chosen randomly.
In case of a limit order, the price is also chosen randomly.
Both price and size of neighbor's order are chosen from the same distribution as source orders, but are independent of the source order.
Finally, the neighbor shares his decision with his neighbors, excluding the source neighbor.
His neighbors treat the incoming information in the same way as they would treat the information coming from a source order.
As a result, a single source action can produce a set of many interaction-related actions.
Note that while information cannot be spread back immediately, loops are possible.
Moreover, a single agent can technically produce multiple orders in response to the same source action.

\begin{figure}[h!]
\centering
\includegraphics[width=0.30\textwidth]{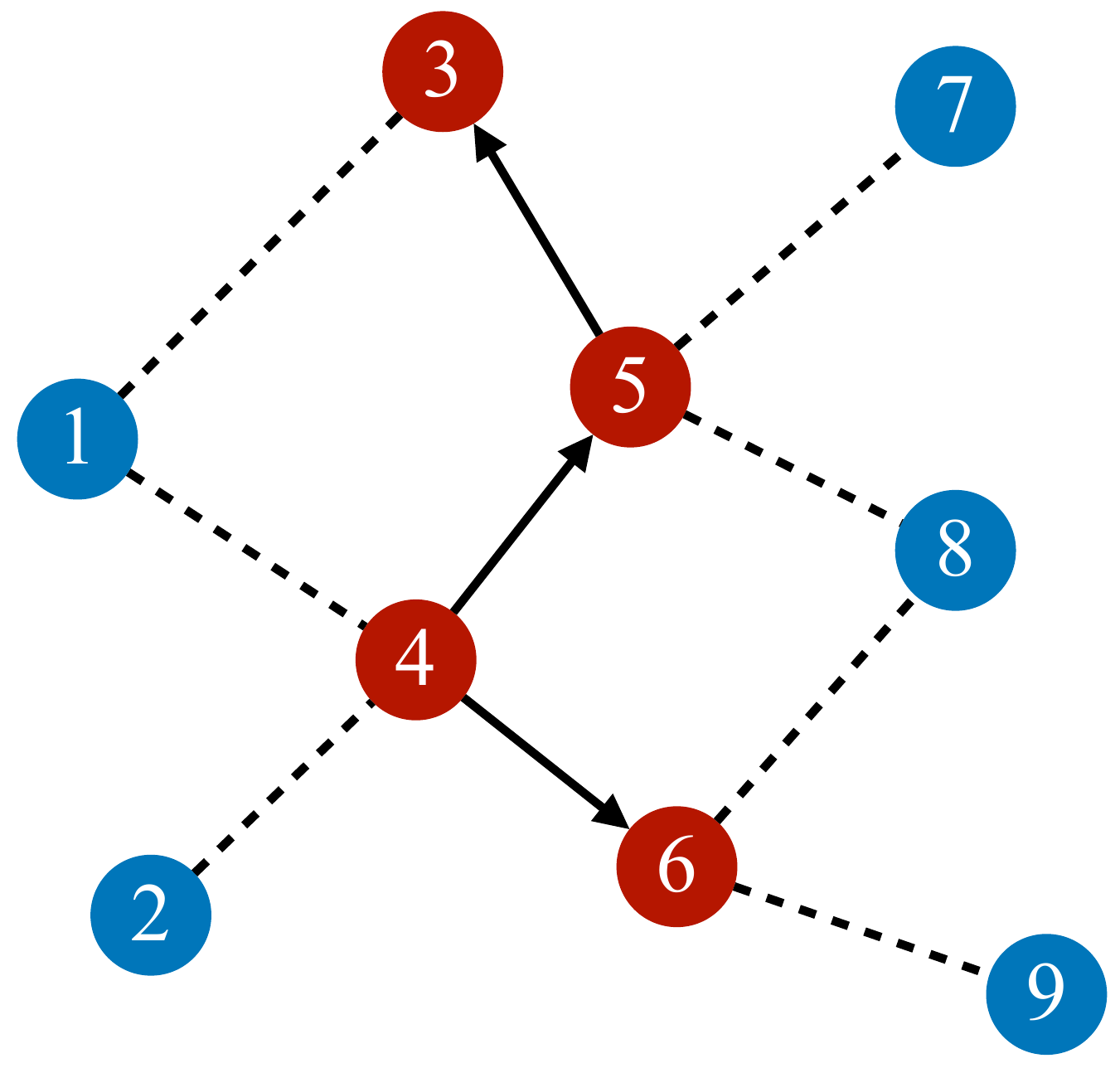}
\caption{A visualization of the exemplary spread described in the text.
Agent number 4 is the source of initial order, whose direction is followed by agents 5 and 6.
Finally, agent 5 spreads the same direction to agent number 3.
\label{fig:example}}
\end{figure}

For clarity, let us go through a simple example of a realization of the presented spreading model.
Imagine that we have nine trading agents numbered from 1 to 9, connected by a network shown in Fig. \ref{fig:example}.
The spreading starts with agent number 4 placing a source order.
This information is followed by two out of his four neighbors.
Agent number 6 does not spread the information further, while agent number 5 is followed by agent number 3.
This dynamics is highlighted in Fig. \ref{fig:example} by connections being drawn with dashed lines (no spreading along this edge), and arrows (there was a spread of information in the direction of the arrow).
All the orders in the spread are of the same type and have the same direction, while their volume (and price, in case of a limit order) is chosen randomly and independently.
Let us denote the time when an order is placed within this spreading by agent $i$ with $\tau^i$.
Then $\tau^5 - \tau^4$, $\tau^6 - \tau^4$, and $\tau^3 - \tau^5$, are all exponentially distributed.
Note that it would be possible to have, for example, agent number 6 placing two follow-up orders.
This could happen if, for example, agent number 6 is followed by agent number 8, and then the latter spreads his decision to agent number 6.

\section{Results}\label{sec:results}

Unless stated otherwise, all the results are obtained by simulating $N = 1000$ agents on a network with $M = 4000$ connections, and with spreading probability $q = \frac{N}{4M}$.
The tick size of the price is equal to $0.01$.
Each agent is parametrized by $\lambda_l = 5000$, $\lambda_l = 20000$, $\lambda_c = 40000$, $\lambda_f = 1000$, $m_s = 5$, $d_s = 1.5$, and $d_p = 2$.
Note that this way, interaction related activities operate in a smaller timescale than individual source actions.
For each scenario, we generate five realizations of the dynamics, which lasts for $720000$ time units.
Each realization is generated using a different random seed and a different underlying network (if applicable).
Only the data after the first $72000$ time units is used to compute the results, to make sure that there is no bias related to the initial stabilization of the dynamics.
In what follows, we look at four statistical properties of the generated limit order book dynamics: (i) the distribution of price returns, (ii) the autocorrelation of absolute price returns, (iii) the autocorrelation of consecutive trades' directions, and (iv) the distribution of inter-event times.
The price returns are defined as $r_t = \log p_t - \log p_{t-1}$.
The autocorrelation of absolute price returns is defined as:
\begin{equation}
    \rho_r(\tau) = \frac{\sum_t (r_{t+\tau} - \bar{r})(r_t - \bar{r})}{\sum_t (r_t - \bar{r})^2},
\end{equation}
where $\bar{r}$ is the average return.
The autocorrelation of consecutive trades' directions is defined as:
\begin{equation}
    \rho_s(k) = \frac{\sum_i (s_{i+k} - \bar{s})(s_i - \bar{s})}{\sum_i (s_i - \bar{s})^2},
\end{equation}
where $s_i$ is the direction of the $i$th trade (it is equal to $1$ for bid side, and $-1$ for ask side), and $\bar{s}$ is the average direction over all trades.
Note that while $\rho_r(\tau)$ is computed in real time, $\rho_s(k)$ is based on event time.
This means that the distance of two consecutive trades is always equal to $1$, regardless of the difference between their timestamps.
Finally, inter-event times are simply defined as $\delta t_i = t_i - t_{i-1}$, where $t_i$ is the timestamp of the $i$th event in the market.
By event we mean any action performed by any agent, this means either posting an order, or canceling one.
All four of the selected properties are related to some well-known \textit{stylized facts} observed in financial data \cite{bouchaud2018agent}.

We use four basic scenarios to test how different connectivity patterns affect market dynamics.
We start with no interaction case, which will serve as a benchmark.
In this case agents' behavior is limited to what was described in Section \ref{sec:no-interaction}.
Then we move to three distinct models of networks, for which we fix the average degree to $\langle k \rangle = 8$.
First case is a modified square lattice lattice, where extra connections are added.
To be more specific, each square of connections is completed by two connections forming the square's diagonals.
This way each node has eight neighbors instead of four.
The resulting graph has high clustering coefficient (many triangles), fixed degree for all nodes, and high average distance between nodes (proportional to $\sqrt{N}$).
Second case is a standard random graph model known as the Erd\H{o}s-R\'{e}nyi model \cite{bollobas2011random}.
In this case $M$ out of $\frac{N(N-1)}{2}$ possible edges are chosen randomly, using a uniform distribution.
Such graphs are characterized by low clustering, Poissonian degree distribution, and low average distance between nodes (proportional to $\log{N}$).
Finally, we use the Barab\'{a}si-Albert model to generate a scale-free network \cite{barabasi1999emergence}.
The resulting graph is characterized by a power-law degree distribution, which produces large degree hubs, and short paths.
The three types of networks present a diverse set of properties.
Now we test how these different properties affect the market dynamics.

\begin{figure}[htb]
  \centering
  \begin{subfigure}[t]{0.49\textwidth}
    \centering
    \includegraphics[width=\linewidth]{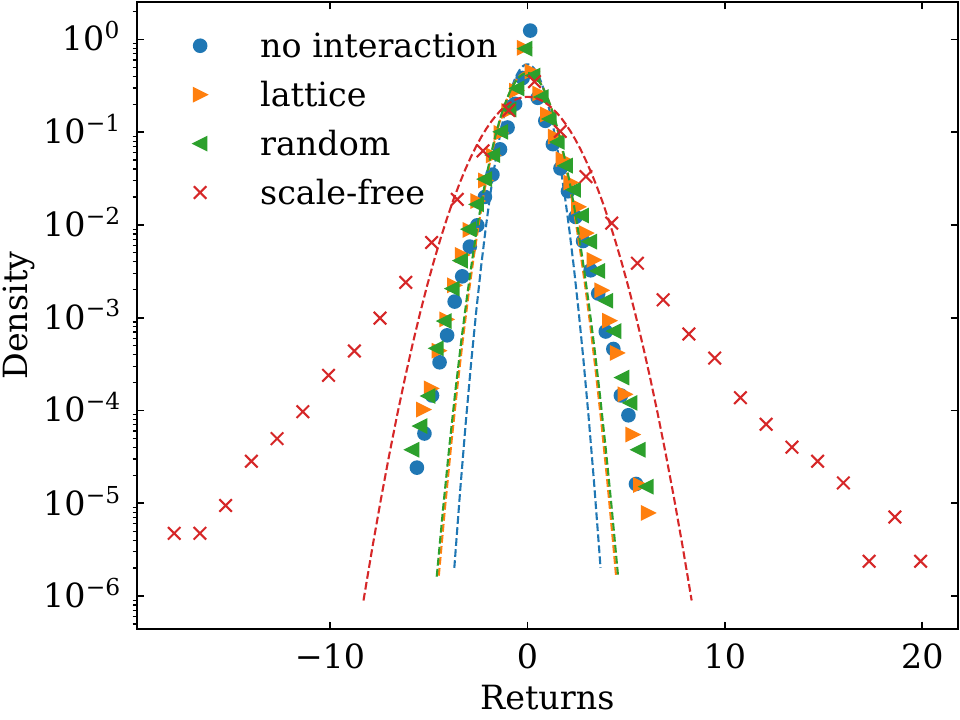}
    \caption{The distribution of returns.}
    \label{fig:rets}
  \end{subfigure}\hfill
  \begin{subfigure}[t]{0.49\textwidth}
    \centering
    \includegraphics[width=\linewidth]{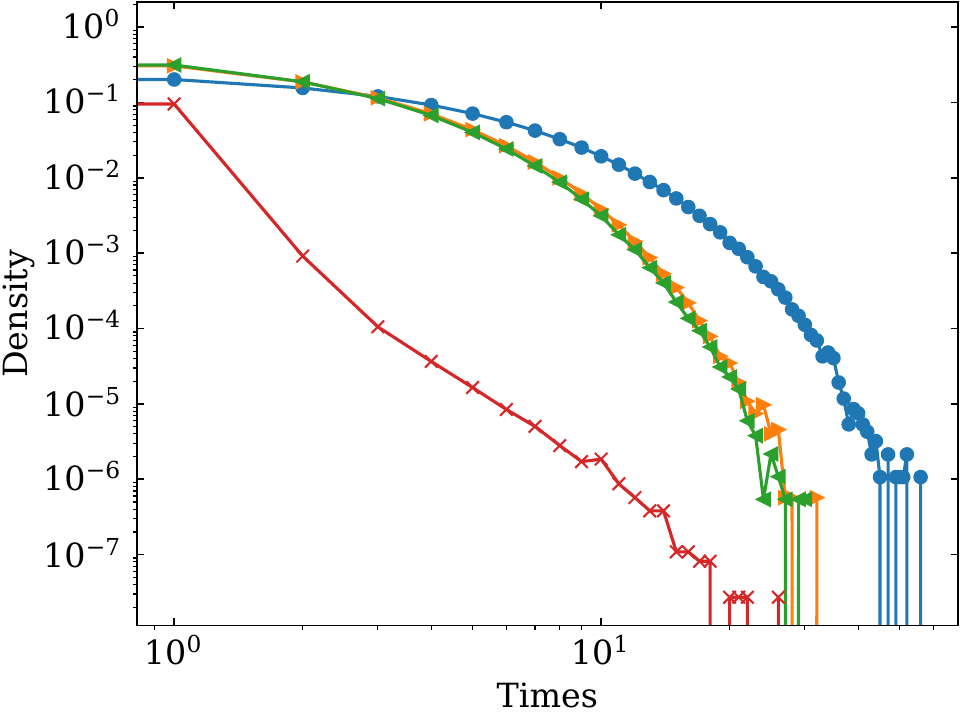}
    \caption{The distribution of waiting times.}
    \label{fig:times}
  \end{subfigure}

  \vspace{5mm} 

  \begin{subfigure}[t]{0.49\textwidth}
    \centering
    \includegraphics[width=\linewidth]{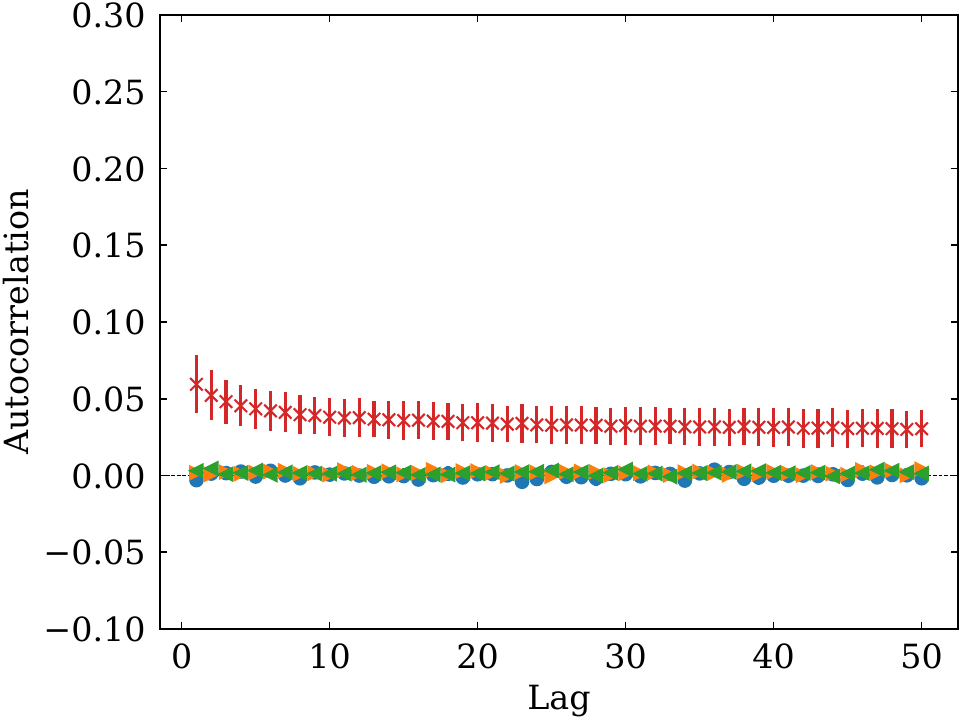}
    \caption{The autocorrelation of trades' signs.}
    \label{fig:sign_acorr}
  \end{subfigure}\hfill
  \begin{subfigure}[t]{0.49\textwidth}
    \centering
    \includegraphics[width=\linewidth]{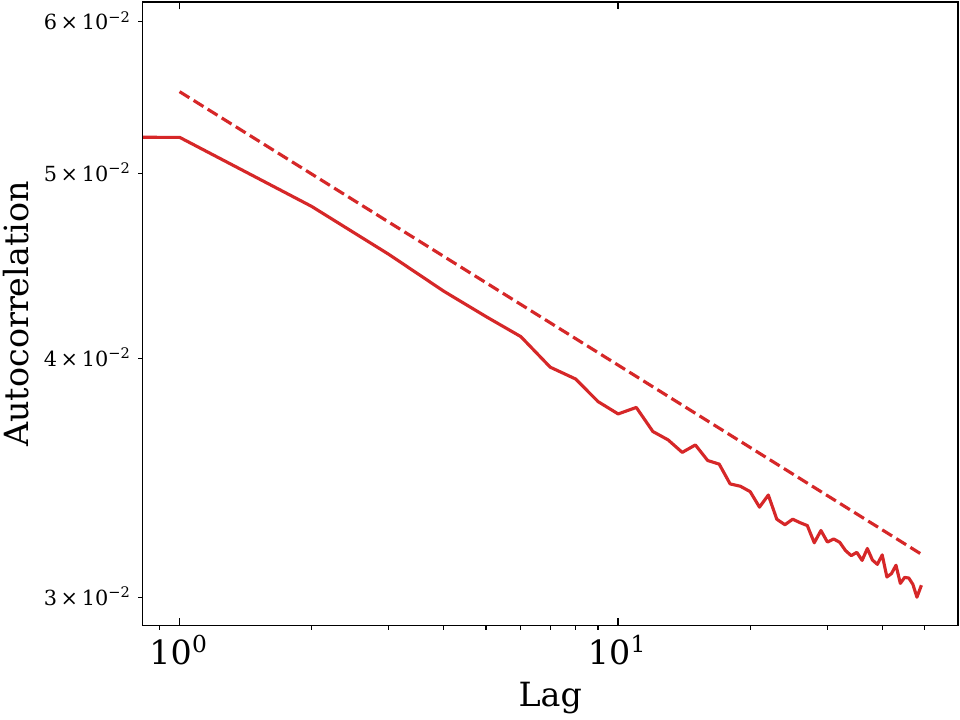}
    \caption{Same as (c) but with $\log\log$ scale.}
    \label{fig:fit_sign_acorr}
  \end{subfigure}

  \vspace{5mm} 

  \begin{subfigure}[t]{0.49\textwidth}
    \centering
    \includegraphics[width=\linewidth]{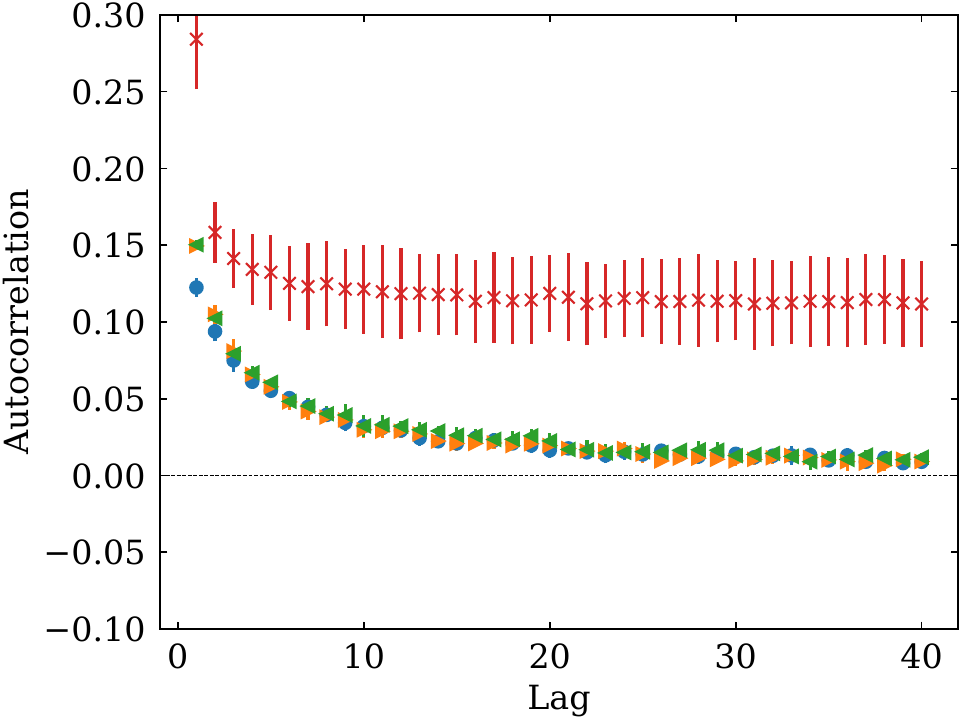}
    \caption{The autocorrelation of absolute returns.}
    \label{fig:abs_ret_acorr}
  \end{subfigure}\hfill
  \begin{subfigure}[t]{0.49\textwidth}
    \centering
    \includegraphics[width=\linewidth]{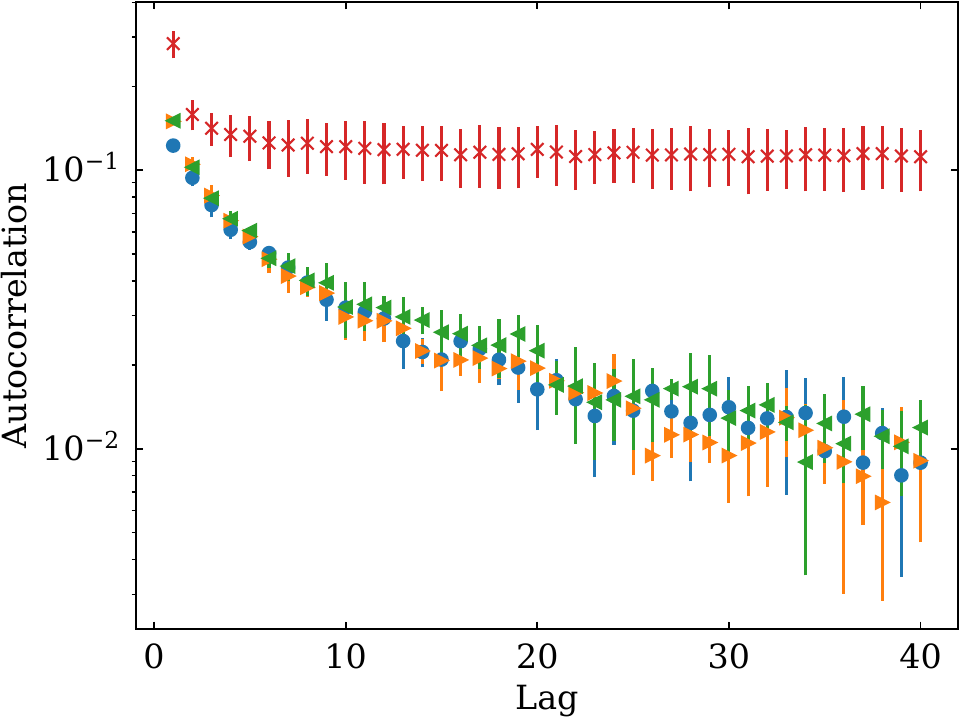}
    \caption{Same as (e) but with $\log$ y-scale}
    \label{fig:log_abs_ret_acorr}
  \end{subfigure}

  \caption{Different statistical properties of four scenarios of simulations: (i) no interaction case, where no network is used, (ii) extendeed square lattice with additional, regular connections, (iii) random graph generated with the Erd\H{o}s-R\'{e}nyi model, and (iv) scale-free graph generated with the Barab\'{a}si-Albert model.
  In all cases $N = 1000$ agents are used with parameters set to: $\lambda_l = 5000$, $\lambda_l = 20000$, $\lambda_c = 40000$, $\lambda_f = 1000$, $m_s = 5$, $d_s = 1.5$, and $d_p = 2$.
  The tick size in the simulation is equal to $0.01$ price unit.
  Each network has $M = 4000$ edges.
  Results are either a sum over five different realizations of simulations procedure (distribution of returns and distribution of waiting times), or an average over them (all autocorrelation results).
  Different realizations differ also in terms of used network (where applicable).}
  \label{fig:results}
\end{figure}

One of the most recognized stylized facts about the price movement is that the price returns have heavy-tailed, non-Gaussian distribution.
In Fig. \ref{fig:rets} the distribution of returns for the different scenarios is shown, together with Gaussian fits drawn with dashed lines.
For the no interaction scenario, the returns have a bit broader tails than the variance would suggest, as we see by comparing them to the dashed blue line representing the corresponding Gaussian distribution.
Adding a lattice or random connections, increases the variance, but the tails stay the same.
It is only scale-free network scenario, where we see a qualitative difference.
The tails produced in this last scenario are not only broader, but as we clearly see by using a $\log$ y-scale, they have a different shape.
More importantly, this shape resembles what is observed for real market returns.

Time dimension plays an important role in trading and risk management.
Similarly to price returns, waiting times (between market events) are also distributed in a way which increases uncertainty and risk.
This is not reflected by the results obtained without interaction, or with lattice and random graphs connecting agents.
As shown in Fig. \ref{fig:times}, only scale-free connectivity is able to reshape the distribution of waiting times.
While the no interaction, lattice, and random graph scenarios produce exponential tails, scale-free network scenario is much closer to the expected power-law distribution.

Real market price dynamics is difficult to predict, due to high market efficiency.
Nonetheless, there are long memory effects observed when analyzing market activities.
One example is the autocorrelation of trades' signs, which has long, power-law tail.
In our experiments, this property, as shown in Fig. \ref{fig:sign_acorr}, is only reproduced by scale-free network scenario.
All other cases produce no autocorrelation in this domain.
Moreover, a linear fit shown in Fig. \ref{fig:fit_sign_acorr} suggests close to power-law behavior for scale-free networks.
Memory effects in financial data are also observed when computing the autocorrelation of absolute price returns.
When using the no interaction model, we observe only short memory, with autocorrelation tail quickly approaching zero, as seen in Fig. \ref{fig:abs_ret_acorr} and \ref{fig:log_abs_ret_acorr}.
Once network is added, in a form of a lattice or a random graph, we see a small increase in short term, but no effect on the tail.
However, if the interaction is modeled with a scale-free network, the effect looks more like what we expect from real market data.
It needs to be pointed out though, that red curves in Fig. \ref{fig:sign_acorr} and \ref{fig:fit_sign_acorr}, while having a promising shape, present lower values of correlation than seen in empirical data.

\section{Conclusions and Discussion}\label{sec:discussion}

To the best of our knowledge, this work is the first attempt to model exogenous interactions between trading agents, using a spreading process on a network, coupled with zero-intelligence investors.
By exogenous we mean that these interactions are not mediated through the limit order book, and are independent from the market state.
This way we investigate how the interaction network structure impacts statistical properties of the limit order book dynamics.
We focus on four classical stylized facts, and test them against four different connectivity scenarios.

The main result is that interaction paired with scale-free connectivity can significantly change the statistical properties of market dynamics.
Importantly, it reproduces known stylized facts, while no interaction scenario fails to do so.
High clustering present in regular lattice, or short diameter generated by Erd\H{o}s-R\'{e}nyi model, did not produce results significantly different from the no interaction case.
Taking into account the types of networks used in the experiments, large spreading hubs seem to drive the observed statistical properties.

While interaction mechanism combined with scale-free connectivity reproduces stylized facts observed in the market, we do not expect it to be the only or even the main origin of long memory, heavy tails etc.
Markets are complex and adaptive systems, driven by many different forces and mechanisms.
It is, however, important to understand, which mechanisms lead to certain market properties.
For example, as suggested in \cite{umar2021tale}, and confirmed by our results, exogenous interactions, driven by scale-free connectivity, can increase risk and produce more outliers in market behavior.

The reason why scale-free networks are capable of qualitative change in the observed statistical properties is the existence of the hubs.
Every time information is spread to one of the highly connected agents, we observe a \textit{burstiness} in activity.
As shown in Section \ref{sec:results}, despite each spreading having exponential waiting time, the resulting distribution is far from it.
In some way, hubs have similar effect here as self-excitation in Hawkes process \cite{abergel2015long}.
However, in our model, the mechanism is not just put into the model, it is simply a result of using a specific network structure.
We believe that this makes the model much more elegant and informative.

In the proposed model there are several restrictive assumptions, including a static network, fixed spreading probability, independence from profitability, and more.
Real process is clearly more complex, but the simplified model gives us insight about what may happen even if the underlying mechanism is different.
If, for example, interaction network is a full graph, but with heterogeneous spreading probability, highly connected agents (in terms of probability strength, rather than degree), could have similar effect as hubs in scale-free networks.
One could also try to predict results of dynamic networks by looking at their snapshots' structure.
In principle, even if the interaction is mostly mediated by the market, a connection can be made.
Imagine agents following actions observed through the market, but with a mechanism giving certain actions higher or lower connectivity (due to timing or other constraints).
This could be modeled with a bipartite graph of agents and events.
Then interaction graphs proposed in this work, would be interpreted as a collapse of the bipartite structure.

Results in this work present only an initial attempt to understand the interplay between market forces and external interactions.
There are multiple directions, which can be followed in the future.
As suggested above, there are many mechanisms, which could be similar in nature to what we proposed.
Network science developed multiple complex extensions to the classical spreading models \cite{ji2023signal}.
Higher-order interactions, multilayer networks, co-evolving networks, all these and more could be used to extend the proposed agent-based model of the market.
On the other hand, agents need not to be limited to noise traders.
Heterogeneous agents, which also reproduce some stylized facts \cite{staccioli2021agent}, can be mixed or extended with interaction mechanism.
This would allow for a more detailed comparison with the models available in the literature.
Constraining agents budget could also have significant impact on memory and synchronization.
Finally, we want to understand how the results depend on the propagation probability.
More specifically, how the so-called \textit{epidemic threshold} \cite{pastor2015epidemic} affects the results.
This is particularly interesting in light of the fact that the threshold for scale-free networks diverges to zero in the thermodynamic limit.

\section*{Acknowledgements}

M. Wilinski acknowledges support from the European Union’s Horizon Europe programme under the Marie Skłodowska-Curie Actions (Grant Agreement No. 101066936, Project: DataABM).

\bibliographystyle{plain}
\bibliography{refs}

\end{document}